\documentclass[11pt]{article}
\usepackage[a4paper,margin=23mm,includefoot]{geometry}
\usepackage[T1]{fontenc}
\usepackage[utf8]{inputenc}
\usepackage{lmodern}
\usepackage{microtype}
\usepackage{amsmath,amssymb,amsthm}
\usepackage{booktabs,array}
\usepackage{graphicx}
\usepackage{algorithm,algpseudocode}
\usepackage[numbers,sort&compress]{natbib}
\usepackage{xcolor}
\usepackage[colorlinks=true,linkcolor=black,citecolor=black,urlcolor=blue]{hyperref}
\providecommand{\doi}[1]{\href{https://doi.org/#1}{doi:\nolinkurl{#1}}}
\usepackage{caption}
\usepackage{needspace}
\usepackage{float}
\usepackage{titlesec}
\titlespacing*{\section}{0pt}{2.2ex plus .5ex minus .2ex}{.85ex}
\titlespacing*{\subsection}{0pt}{1.65ex plus .4ex minus .2ex}{.55ex}
\hypersetup{pdfauthor={Azzeddine Ihsine and Sara Ihsine},
  pdftitle={Optimal Development Path: An Empirical Evaluation of Tiered Dependency-Aware Sequencing},
  pdfsubject={Dependency-aware sequencing and TAWOS evidence}}

\newtheorem{proposition}{Proposition}

\newcommand{\BVS}{\mathrm{BVS}}
\newcommand{\ERS}{\mathrm{ERS}}
\newcommand{\PVS}{\mathrm{PVS}}

\title{\textbf{Optimal Development Path:}\\
An Empirical Evaluation of Tiered Dependency-Aware Sequencing}
\author{Azzeddine Ihsine \quad Sara Ihsine\\[2pt]
\small Inovionix\\[2pt]
\small \href{mailto:azzeddine.ihsine@inovionix.com}{\texttt{azzeddine.ihsine@inovionix.com}}, \href{mailto:sara.ihsine@inovionix.com}{\texttt{sara.ihsine@inovionix.com}}}
\date{}
\begin{document}
\maketitle
\begin{abstract}
The Optimal Development Path (ODP) in D-POAF orders Feature Blocks in a Wave Candidate by choosing the currently eligible block with the lowest Wave-pinned ERS/BVS ratio. This transparent, dependency-valid rule does not explicitly account for the value unlocked downstream. We compare native ODP (L0) with immediate unlock count (L1), newly unlocked value (L2), discounted multi-hop potential (L3) and a one-step rollout (L4), alongside local search, Sidney decomposition and an exact reference. On held-out synthetic graphs with about one prerequisite per block, L0 reaches 0.978 of the surrogate optimum at 12 blocks, compared with 0.993 for L2 and 0.994 for L4. At 100 blocks, L0 reaches 0.942 and L2 0.983. However, 373 Jira sprints from 21 open-source projects (TAWOS), used only as proxy scheduling instances rather than governed D-POAF Waves, have sparser recorded dependencies (0.12 per issue). On these instances, L0 reaches 0.998 of the surrogate optimum and is exactly optimal in 69\% of sprints. Unlock bonuses tuned on synthetic graphs show no significant average gain on the sprint-derived graphs and worsen the surrogate in 44--59\% of sprints. L4 improves it by about 0.8 thousandths and is guaranteed not to worsen that \emph{surrogate}, not operational outcomes. Counterfactual calendar replays using Jira time-in-status records, with missing durations imputed, find that improving the surrogate does not reliably improve value-weighted completion. The historical issue-resolution order reaches 0.929 of the surrogate optimum but does not reveal the order in which work started. These findings support keeping L0 as the default and presenting rollout and exact references as auditable, read-only alternatives, while treating unlock bonuses as context-dependent research options rather than production defaults. Valid sequences make declared prerequisite readiness explicit, but the study does not measure whether this prevents interruptions or rework during execution.
\end{abstract}


\section{Introduction}
Prioritizing software work is not only a matter of assigning scores to requirements. Teams must decide which needs merit limited resources, what can be developed in the presence of prerequisites, and how to revise an ordering when the available evidence changes. Value-based software engineering and cost-value analysis established business value and development cost as explicit decision criteria~\cite{boehm2003vbse,karlsson1997cost}. Subsequent research proposed numerous approaches to requirements prioritization, yet empirical support and applicability vary across settings. A review of 102 studies reported that many proposed techniques had not been evaluated in real-life contexts~\cite{bukhsh2020slr}. More recent reviews document both the range of task-ranking methods and continuing concerns about scale and practical evaluation~\cite{bugayenko2023taskreview,yaseen2025scalability}. An industrial study further found that the criteria influencing prioritization decisions change as requirements move through development~\cite{svensson2024criteria}. These findings argue against treating any one scoring rule as a universally appropriate development order.

Two decisions are often discussed together but must be separated analytically. The Next Release Problem addresses \emph{selection}: which requirements should enter a release under resource and dependency constraints~\cite{bagnall2001nrp,pitangueira2015sbse}. Once a set has been selected, \emph{sequencing} asks in which precedence-valid order its constituent items should be considered. Industry-inspired ratios such as Weighted Shortest Job First (WSJF) offer interpretable priorities based on delay cost and size~\cite{reinertsen2009flow,leffingwell2011agile}, but a ratio alone does not establish an optimal order under general precedence relations. The distinction also matters for evaluation. A ranking may perform well on a score-based criterion without improving completion dates or value realized after release.

Operations research provides more demanding comparators than simple priority scores. Smith's rule characterizes the optimum of total weighted completion time when there are no precedence constraints~\cite{smith1956}. General precedence-constrained sequencing is considerably harder~\cite{lawler1978sequencing}, and Sidney's decomposition, followed by later decomposition and approximation work, makes explicit how groups of dependent jobs can alter an otherwise appealing local decision~\cite{sidney1975decomposition,chekuri1999precedence,margot2003decompositions}. Rollout methods offer a different route: evaluate a proposed next action through a complete heuristic continuation, under conditions that can guarantee improvement on the \emph{declared objective}~\cite{bertsekas1997rollout}. These results are relevant to software planning, but their classical processing times and weights do not automatically represent organizational risk, elapsed calendar time, or observed business value.

D-POAF supplies the concrete decision context studied here. A \emph{Wave} is a unit of verifiable progress and a governance container spanning Instruct \& Scope, Shape \& Align, Execute \& Evolve, and Learn \& Adapt, with defined scope, delegation limits and proof obligations (PoD, PoV and PoR)~\cite{dpoaf_spec,dpoaf_terms}. Within a Wave, a candidate set of \emph{Feature Blocks} and their declared prerequisites is evaluated through a versioned Matrix snapshot. The native Optimal Development Path (ODP) then orders currently eligible blocks by the stored effort-and-risk-to-value ratio, subject to dependency closure and deterministic tie-breaking~\cite{dpoaf_guide}. This rule is transparent but greedy: it cannot explicitly credit a lower-ranked prerequisite for high-value blocks it unlocks later. Our companion study reported a measurable gap from a specified score-based optimum on synthetic graphs~\cite{ihsine2026governed}. That result motivates further comparison, not an assumption that a more elaborate algorithm will improve delivery.

For empirical comparison, publicly available Jira histories are useful but imperfect. TAWOS provides sprint issue sets, effort proxies, links and status records suitable for reconstructing precedence graphs~\cite{tawosi2022tawos,tawosdata}. A Scrum sprint is a timeboxed event, \emph{not} a D-POAF Wave or a substitute for Wave governance and evidence obligations~\cite{scrumguide2020}. In this paper, sprint-derived issues are therefore \emph{proxy scheduling items}, their reported priorities and story points are \emph{proxy inputs}, and the evaluation tests the transfer of sequencing behaviour to observed graph structures. Neither those records nor retrospective calendar replays establish that a D-POAF Wave was governed, that a Feature Block's business value was delivered, or that production costs fell.

The research question is consequently specific: \emph{when does explicit anticipation of downstream prerequisites improve the native ODP ordering on a declared surrogate objective, and how well do such improvements transfer across graph structures and imperfect duration proxies?} We retain native ODP as Level~0 and compare four anticipation strategies with established scheduling baselines. The contributions are:
\begin{enumerate}
\item A formal statement of the Wave-pinned ODP implementation contract (closure, eligibility, distance, PVS tie-break, names, stored rounding) and a measurement of its tie-break and rounding effects.
\item A transparent family of unlock-based priorities (L1--L3) and a one-step rollout (L4) whose non-worsening property follows from rollout theory.
\item An exact reference for instances whose Sidney modules fit within a dynamic-programming limit, validated against brute-force dynamic programming on small instances.
\item A synthetic benchmark harmonized with the companion article, with strength parameters selected on training seeds and evaluated on held-out seeds, and exact optima up to 100 blocks.
\item An evaluation on 373 historical Jira sprints from 21 open-source projects, transformed into scheduling proxies rather than treated as D-POAF Waves, with exact optima, retrospective calendar replays and recorded issue-resolution orders.
\item A governed adoption design, in which alternatives are shown as read-only comparisons with their surrogate and calendar evidence, plus a distinct acceptance case from the Blackridge Atlas private beta. This case examines a configured Wave graph, not operational cost or value outcomes.
\end{enumerate}

\section{Related Work}
The literature on prioritization covers cost-value ranking, release selection, task and issue ranking, and multi-criteria decision support~\cite{karlsson1997cost,bagnall2001nrp,bukhsh2020slr,bugayenko2023taskreview}. These tasks do not share an identical decision variable. In particular, the Next Release Problem chooses a subset under a budget, whereas the present study sequences a fixed selected set with prerequisites. The practical WSJF-style baseline used here divides a business-value proxy by effort. This is \emph{not} a faithful reconstruction of a full SAFe prioritization decision, whose Cost of Delay requires information that Jira priorities do not supply~\cite{reinertsen2009flow,leffingwell2011agile}.

For our specified surrogate, unconstrained ordering is governed by Smith's rule~\cite{smith1956}. With general precedences the corresponding problem $1|\mathrm{prec}|\sum w_jC_j$ is strongly NP-hard~\cite{lawler1978sequencing}. Sidney's decomposition extracts precedence-closed initial sets of maximal weight-to-processing-time ratio. Subsequent work relates these decompositions to approximation guarantees and computational formulations~\cite{sidney1975decomposition,chekuri1999precedence,margot2003decompositions,correa2005single}. Computing maximal-closure sets can be reduced to network flow, with fractional optimization providing the ratio search~\cite{picard1976closure,dinkelbach1967fractional}. Separately, one-step rollout can improve a sequentially consistent base heuristic on the \emph{same} objective used to evaluate its continuations~\cite{bertsekas1997rollout}. The present study treats these results as established methods and asks how closely constrained D-POAF score-based sequences approach them.

Empirical transfer requires equal care. Reviews show that experimental demonstrations do not necessarily establish practical suitability~\cite{bukhsh2020slr,yaseen2025scalability}, while industrial evidence indicates that relevant prioritization criteria depend on the decision stage~\cite{svensson2024criteria}. TAWOS exposes data for comparative analysis~\cite{tawosi2022tawos}, but neither recorded Jira priority nor story points recover a governed BVS or ERS. Our retrospective replays therefore probe sensitivity to real graph structures and duration proxies, not end-to-end D-POAF governance, cost savings, or observed post-release value.

\section{Wave Candidate Model and Native ODP}
For every selected Feature Block $i$, a frozen Matrix snapshot stores value $\BVS_i>0$, a composite effort-and-risk score $\ERS_i=1+\beta E_i+\gamma R_i>0$, a priority $\PVS_{\alpha,i}=\BVS_i^\alpha/\ERS_i$, and prerequisites $P(i)$. ODP uses the distance
\begin{equation}
d_i=\frac{\ERS_i}{\BVS_i}. \label{eq:d}
\end{equation}
Ordering a fixed selected set must be distinguished from selecting which blocks enter a budget-constrained release. The latter is studied in the companion article~\cite{ihsine2026governed}.

\paragraph{Surrogate objective.} For a dependency-valid order $\pi$ of a fixed set, define the ERS completion burden and the objective
\begin{equation}
C_i^{\ERS}(\pi)=\sum_{j:\,\pi(j)\le\pi(i)}\ERS_j,\qquad J(\pi)=\sum_i\BVS_i\,C_i^{\ERS}(\pi). \label{eq:J}
\end{equation}
Lower $J$ means that high-value blocks become available earlier under cumulative effort-and-risk burden. For adjacent unconstrained blocks, $J_{ij}-J_{ji}=\BVS_j\ERS_i-\BVS_i\ERS_j$, so $i$ should precede $j$ exactly when $d_i\le d_j$: without prerequisites, ascending $d$ is optimal for Eq.~\eqref{eq:J} (Smith's rule~\cite{smith1956}). This is a theorem for the declared surrogate, not evidence of shorter schedules. $J$ is not calendar time: ERS is a governed score, not an observed duration (Section~\ref{sec:rq3}).

\paragraph{Implementation contract.} Matrix snapshots compute and persist $d_i=\mathrm{round}_{4}(\ERS_i/\BVS_i)$ and $q_i=\mathrm{round}_{2}(\BVS_i^{\alpha_w}/\ERS_i)$ with the exponent $\alpha_w$ pinned to Wave $w$ (default 2). ODP reads these stored values and does not recompute PVS with another exponent. All selected blocks and their transitive prerequisites are included, scores and references are validated, and the next block minimizes the tuple
\begin{equation}
(d_i,\,-q_i,\,\mathrm{moduleName}_i,\,\mathrm{blockName}_i)\quad\text{over}\ \{i\notin D: P(i)\subseteq D\}, \label{eq:tuple}
\end{equation}
where $D$ is the set already placed (Algorithm~\ref{alg:odp}). For positive scores, $\PVS_{1,i}=1/d_i$, so for $\alpha_w=1$ the PVS tie-break is redundant before rounding. For $\alpha_w=2$ it favours the higher BVS among blocks of equal distance. The rule is neither a critical-path analysis nor an explicit valuation of downstream unlocking.

\begin{algorithm}[t]
\caption{Native ODP (Level 0), pinned Matrix snapshot}\label{alg:odp}
\begin{algorithmic}[1]
\Require Targets $T$, snapshot $(d_i,q_i,P(i),\mathrm{module}_i,\mathrm{block}_i)$
\State $S\gets \textsc{TransitiveDependencyClosure}(T)$, then validate references, positive scores and snapshot consistency
\State $D\gets\emptyset$, $\pi\gets[\,]$
\While{$|D|<|S|$}
  \State $A\gets\{i\in S\setminus D: P(i)\subseteq D\}$
  \If{$A=\emptyset$} \Return \textsc{Infeasible} \Comment{cycle or unresolved prerequisite}\EndIf
  \State $i^\star\gets\arg\min_{i\in A}(d_i,-q_i,\mathrm{module}_i,\mathrm{block}_i)$ \Comment{lexicographic}
  \State append $i^\star$ to $\pi$, then set $D\gets D\cup\{i^\star\}$
\EndWhile
\State \Return $\pi$
\end{algorithmic}
\end{algorithm}

\section{Anticipation Levels}\label{sec:levels}
Levels L1--L3 change only the preference among currently eligible blocks. For the placed set $D$ and an eligible candidate $i$, let $U(i,D)=\{k\notin D\cup\{i\}: P(k)\subseteq D\cup\{i\},\ i\in P(k)\}$ be the blocks that become eligible exactly when $i$ is placed, and let $h(i,k)$ be the number of arcs on a shortest path from $i$ to a descendant $k$. The downstream credits are
\begin{align}
g_c(i,D)&=|U(i,D)|, & g_v(i,D)&=\sum_{k\in U(i,D)}\BVS_k, & g_h(i,D)&=\sum_{k\in\mathrm{desc}(i)\setminus D}\BVS_k\,\delta^{\,h(i,k)-1}, \label{eq:g}
\end{align}
with $\delta=0.5$, and the priority of level $k\in\{c,v,h\}$ (L1, L2, L3) is
\begin{equation}
S_k(i,D,\theta)=\frac{\BVS_i+\theta\,g_k(i,D)}{\ERS_i}, \label{eq:S}
\end{equation}
highest first, ties broken by the L0 tuple~\eqref{eq:tuple}. For $\theta=0$ every level reproduces L0. A bonus counts the value of future blocks that will themselves be scored later, so double counting is a real risk. Moreover, $g_c$ and $g_h$ are in different units, so a shared $\theta$ does not mean a shared economic weight.

\paragraph{Level 4: one-step rollout.} At each step, every eligible block $i$ is tried as next block, the remaining blocks are ordered by L0, and the candidate whose complete sequence has the lowest $J$ is kept (ties by the L0 tuple). Rollout uses only the stored scores and prerequisites and has no tuned parameter.

\begin{proposition}\label{prop:rollout}
$J(\pi_{L4})\le J(\pi_{L0})$ on every instance.
\end{proposition}
L0 is sequentially consistent: the L0 order of the blocks remaining after any L0 prefix is the remainder of the L0 order. Rollout with a sequentially consistent base heuristic never does worse than the heuristic~\cite{bertsekas1997rollout}. The guarantee concerns the surrogate~\eqref{eq:J} only.

\section{Exact Reference Beyond Twelve Blocks}\label{sec:exact}
Subset dynamic programming limits exact optima to about 12--16 blocks. For larger graphs we use the structure of the problem: every optimal schedule for Eq.~\eqref{eq:J} is consistent with a Sidney decomposition into successive ratio-maximal initial sets~\cite{sidney1975decomposition,chekuri1999precedence,margot2003decompositions}, so the optimum is the concatenation of optimal orders of the successive modules. We compute each module as an inclusion-minimal precedence-closed set of maximal $\sum\BVS/\sum\ERS$ (Dinkelbach iterations over maximum-weight closures solved by minimum cut~\cite{picard1976closure,dinkelbach1967fractional}) and order it exactly by subset dynamic programming. The method is exact as long as every module has at most 20--22 blocks. It reproduced the brute-force optimum on 600 random 12-block instances (one third with integer, heavily tied scores) and on the 105 real sprints with at most 16 blocks (three score variants each), with a largest relative difference of $4\times10^{-16}$. On real sprints the largest module had 21 blocks. In the synthetic generator at $\lambda=1$ the largest modules had 12, 18 and 20 blocks at $n=25$, 50 and 100. Eleven of 150 instances at $n=100$ exceeded the limit and are excluded. Two reference heuristics are also reported: \emph{ODP + local search} (first-improvement insertion moves that preserve validity, started from L0) and \emph{Sidney (ODP in modules)}, the decomposition with L0 inside each module.

\section{Research Questions and Methods}
\begin{description}
\item[RQ1] On synthetic graphs, how far is L0 from the exact optimum of $J$, and how much do L1--L4 recover, from 12 to 100 blocks?
\item[RQ2] Do these algorithmic results transfer to precedence graphs reconstructed from historical Jira sprints, without treating those sprints as D-POAF Waves?
\item[RQ3] Do gains on the surrogate carry over to value-weighted completion in calendar time with durations the policies do not see?
\item[RQ4] What do the pinned exponent, stored rounding, a surrogate-only gate and the BA-W1 private-beta acceptance case reveal for governed adoption?
\end{description}

\subsection{Synthetic benchmark}\label{sec:m-synth}
The generator is that of the companion article~\cite{ihsine2026governed}: $\BVS\sim N(5.5,2^2)$ truncated to $[1,10]$, effort lognormal (median 5, log-s.d.\ 0.7, truncated to $[1,40]$), risk $U[1,10]$, $\ERS=1+0.1E+0.1R$, and $\min(\mathrm{Poisson}(\lambda),3)$ prerequisites among earlier blocks. At $n=12$ we use $\lambda\in\{0,0.5,1,1.5\}$ with eight instances per seed and level and the same random keys as the companion article, so L0 values coincide in both papers. The exact value $J^\star$ comes from subset dynamic programming. At $\lambda=1$ graphs carry 0.87 prerequisites per block. The strength $\theta\in\{0.1,0.25,0.5,1\}$ of each of L1--L3 is chosen on seeds 0--14 at $\lambda=1$. All reported $n=12$ numbers come from seeds 15--29. A scaling study uses $n\in\{25,50,100\}$ at $\lambda=1$ (five instances per seed, exact $J^\star$ by Section~\ref{sec:exact}). We use thirty seeds, the random stream \texttt{default\_rng([20260827, study, seed, \ldots])}, and paired seed-level 95\% Student-$t$ intervals.

\subsection{Sprint-derived scheduling proxies (TAWOS)}\label{sec:m-real}
TAWOS is a public dataset of 39 agile open-source projects managed in Jira~\cite{tawosi2022tawos,tawosdata}. Our unit of observation is a historical sprint's issue set, temporarily held fixed to define one scheduling instance. In Scrum, sprint scope may be clarified or renegotiated as work progresses~\cite{scrumguide2020}. We do not reconstruct a D-POAF Wave, and a Jira issue is not assumed to be a Feature Block. We use an issue only as a \emph{task-level proxy} with a score and an observed precedence link.

For each sprint with at least one internal dependency, we retain issues not closed as Duplicate, Won't Fix, Invalid or similar. We interpret the Jira links ``$A$ blocks $B$'' and ``$B$ depends on $A$'' between retained issues as $A$ preceding $B$, dropping two cycle-closing links. Dependencies on issues outside the sprint cannot be represented in this within-sprint instance. Effort $E$ is the story-point estimate (missing or above 40 replaced by the sprint median). Story points are relative estimates, not observed person-days. Risk is not recorded, so $R=5$ for every issue, giving $\ERS=1+0.1E+0.5$. Sensitivity checks use $\beta=0.3$ and $\gamma=0$. We map Jira priority to five levels (Trivial/Lowest $=1$ to Blocker/Highest $=5$) and set $\BVS=2\times$level. This is a priority-based proxy, not a D-POAF business-value assessment. The Jira key supplies the final deterministic name tie-break.

These transformations yield 373 sprint-derived instances from 21 projects (median 25 issues, maximum 195) with 0.12 recorded internal prerequisites per issue on average. L1--L3 use $\theta$ selected on synthetic training seeds, without tuning on these instances. Because sprints within a project are not independent, uncertainty is computed over 21 paired project means. The data can test sequencing behaviour under observed link structures, but cannot test Matrix governance, Wave closure, delegation bounds or proof generation.

\subsection{Calendar criterion}\label{sec:m-cal}
To test an outcome that no policy sees, we replay each order under a hypothetical non-delay list schedule with one or three teams: whenever a team is free, it starts the first block of the order whose prerequisites are finished. We report $\min_{\text{orders}}\sum_i\BVS_iC_i\,/\,\sum_i\BVS_iC_i$, where $C_i$ is the calendar completion time and the minimum runs over all evaluated orders. For these retrospective replays, durations are based on minutes in Jira status ``In Progress''. A total of 313 sprints have at least one such record, the median sprint has it for 67\% of its issues, and missing durations are imputed from the sprint's minutes per story point. On synthetic graphs ($n=25$, $\lambda=1$, eight instances per seed), durations are $E_i\exp(\varepsilon_i)$ with $\varepsilon_i\sim N(0,0.35^2)$. Crediting each block at its completion is a scenario assumption: if a Wave is released atomically, block-level timing overstates the effect of reordering.

\subsection{Governance checks}\label{sec:m-gov}
\emph{Exponent and rounding.} On the 960 synthetic $n=12$ instances and the 373 sprints we compare the contract~\eqref{eq:tuple} with $\alpha_w=2$ and $\alpha_w=1$, and with unrounded scores. \emph{Surrogate gate.} A retrospective gate keeps an alternative order only when it strictly lowers $J$. On real sprints we then count how often the kept alternative is worse than L0 in calendar time. \emph{Blackridge Atlas private-beta case (BA-W1).} Blackridge Atlas is a defined project scenario carried out in the D-POAF platform private beta, concerned with a business-critical risk-scoring API. Its v1.0 brief specifies four modules and 14 Feature Blocks spanning request intake, risk evaluation, decision output and service basics~\cite{blackridgebrief2026}. BA-W1 refers to the project's first Wave (W1), while BA is its internal project prefix. The BA-ODP-01 acceptance study used a fixed configuration of these 14 blocks and 18 declared prerequisite arcs to examine the native ordering and alternative policies. The project's beta setting is real, but the study protocol describes the numerical BVS/ERS configuration as test data, not as effort, risk or value observed in a deployed banking system. We therefore report a \emph{private-beta acceptance case using configured scoring inputs}, distinct from both randomly generated benchmarks and an independently observed industrial outcome study. No claim about production cost, elapsed delivery time or realized business value follows from this case.

\section{Results}
\subsection{RQ1: anticipation helps on dense synthetic graphs}\label{sec:rq1}

\begin{table}[t]
\centering\small
\caption{Synthetic benchmark: mean $J^\star/J$. Left: $n=12$, held-out seeds 15--29, exact $J^\star$ by subset dynamic programming. Right: $\lambda=1$, all seeds, exact $J^\star$ by Sidney decomposition with module dynamic programming (139 of 150 instances at $n=100$). $\theta$ selected on seeds 0--14.}\label{tab:synth}
\begin{tabular}{lcccccc}
\toprule
& \multicolumn{3}{c}{$n=12$, $\lambda=$} & \multicolumn{3}{c}{$\lambda=1$, $n=$}\\
\cmidrule(lr){2-4}\cmidrule(lr){5-7}
Policy & 0.5 & 1.0 & 1.5 & 25 & 50 & 100\\
\midrule
L0 native ODP & 0.9836 & 0.9784 & 0.9806 & 0.9660 & 0.9530 & 0.9418 \\
$\PVS_2$, eligible greedy & 0.9767 & 0.9724 & 0.9744 & 0.9590 & 0.9469 & 0.9333 \\
Value/effort (WSJF) & 0.9554 & 0.9584 & 0.9640 & 0.9500 & 0.9310 & 0.9243 \\
L1 count ($\theta=1$) & 0.9914 & 0.9914 & 0.9897 & 0.9846 & 0.9789 & 0.9766 \\
L2 value ($\theta=0.25$) & 0.9926 & 0.9928 & 0.9921 & 0.9882 & 0.9842 & 0.9831 \\
L3 multi-hop ($\theta=0.1$) & 0.9924 & 0.9923 & 0.9919 & 0.9859 & 0.9828 & 0.9794 \\
L4 one-step rollout & 0.9926 & 0.9943 & 0.9928 & 0.9846 & 0.9803 & 0.9780 \\
ODP + local search & 0.9992 & 0.9994 & 0.9993 & 0.9983 & 0.9957 & 0.9923 \\
Sidney (ODP in modules) & 0.9997 & 0.9991 & 0.9993 & 0.9993 & 0.9995 & 0.9996 \\
Random topological & 0.8411 & 0.8813 & 0.8981 & 0.8525 & 0.8394 & 0.8287 \\ 
\bottomrule
\end{tabular}
\end{table}

Without prerequisites every policy built on ascending $d$ is exact, which only verifies the implementation. With prerequisites, L0 reaches 0.9836, 0.9799 and 0.9802 of the optimum at $\lambda=0.5$, 1 and 1.5 over all seeds, as in the companion article. Training selects $\theta=1$ for L1, $0.25$ for L2 and $0.1$ for L3. On held-out seeds every level improves on L0 at every intensity (Table~\ref{tab:synth}). At $\lambda=1$, the paired gains are 12.9 [6.3, 19.5] (L1), 14.4 [7.7, 21.1] (L2), 13.9 [7.2, 20.5] (L3) and 15.9 [8.9, 22.9] (L4) thousandths of $J^\star$. Local search and Sidney (ODP in modules) do better still (20.9 and 20.7 thousandths). The gap of L0 grows with size, from 3.4\% at $n=25$ to 5.8\% at $n=100$. L2 recovers about 70\% of it (0.983 at $n=100$), and Sidney (ODP in modules) stays within 0.05\% of the optimum at every size. Strength matters: at $\theta=1$, L2 and L3 fall to 0.965 and 0.952 at $\lambda=1$, below L0.

\subsection{RQ2: on sprint-derived graphs, native ODP is almost optimal}\label{sec:rq2}

\begin{table}[t]
\centering\small
\setlength{\tabcolsep}{2.8pt}
\caption{373 TAWOS sprints (21 projects): mean and minimum $J^\star/J$, share of sprints solved exactly, share where the policy is worse than L0 on $J$, paired gain over L0 in thousandths of $J^\star$ (mean over projects, 95\% CI), and value-weighted calendar completion with recorded durations on 1 and 3 teams, relative to the best evaluated order (313 sprints). L1--L3 use the $\theta$ selected on synthetic training seeds. L4 was not run on the four sprints above 120 issues.}\label{tab:tawos}
\begin{tabular}{lccccccc}
\toprule
Policy & $J^\star/J$ & min & exact & worse & gain over L0 ($10^{-3}$) & 1 team & 3 teams\\
\midrule
L0 native ODP & 0.9981 & 0.885 & 69\% & -- & -- & 0.955 & 0.975 \\
$\PVS_2$, eligible greedy & 0.9968 & 0.798 & 62\% & 13\% & -0.68 [-1.52, +0.15] & 0.953 & 0.974 \\
Value/effort (WSJF) & 0.9909 & 0.884 & 50\% & 33\% & -11.90 [-19.62, -4.17] & 0.970 & 0.983 \\
L1 count ($\theta=1$) & 0.9981 & 0.928 & 34\% & 53\% & -0.08 [-0.98, +0.82] & 0.944 & 0.971 \\
L2 value ($\theta=0.25$) & 0.9975 & 0.973 & 29\% & 59\% & -0.60 [-1.73, +0.52] & 0.938 & 0.969 \\
L3 multi-hop ($\theta=0.1$) & 0.9991 & 0.982 & 43\% & 44\% & +0.73 [-0.14, +1.61] & 0.947 & 0.974 \\
L4 one-step rollout & 0.9992 & 0.972 & 75\% & 0\% & +0.77 [+0.10, +1.43] & 0.952 & 0.977 \\
ODP + local search & 1.0000 & 0.999 & 99\% & 0\% & +1.19 [+0.27, +2.11] & 0.957 & 0.979 \\
Sidney (ODP in modules) & 1.0000 & 1.000 & 100\% & 0\% & +1.19 [+0.27, +2.11] & 0.956 & 0.978 \\
Exact optimum & 1.0000 & 1.000 & 100\% & 0\% & +1.19 [+0.27, +2.11] & 0.954 & 0.977 \\
\midrule
Recorded resolution order & 0.9293 & 0.665 & -- & 94\% & -71.37 [-88.17, -54.56] & -- & -- \\ 
\bottomrule
\end{tabular}
\end{table}

\begin{figure}[t]
\centering
\includegraphics[width=0.82\linewidth]{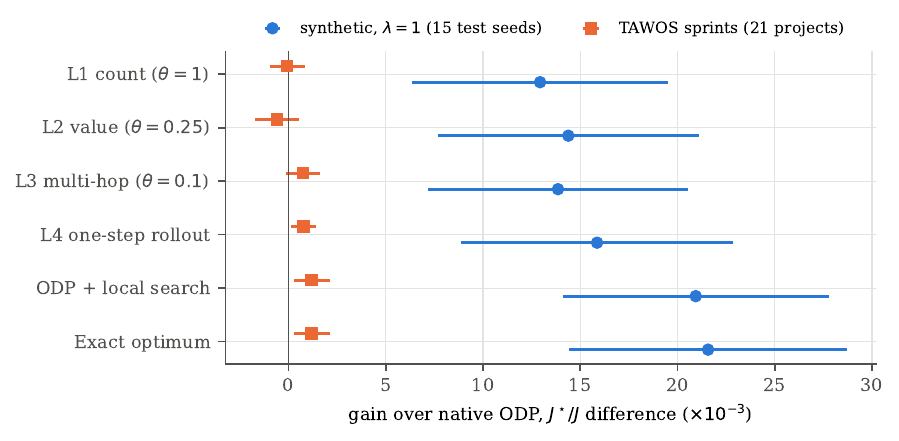}
\caption{Gain over native ODP on the surrogate $J$: synthetic graphs at $\lambda=1$ ($n=12$, held-out seeds) versus real TAWOS sprints (paired project means). Bars are 95\% confidence intervals.}\label{fig:svr}
\end{figure}

\emph{The recorded within-sprint dependency graphs are sparse:} 0.12 prerequisites per issue against 0.87 in the synthetic generator at $\lambda=1$. L0 reaches 0.9981 of the exact optimum on average and is exactly optimal in 69\% of sprints (Table~\ref{tab:tawos}). Among the remaining sprints, the median gap is 0.18\% (maximum 11.5\%). Its efficiency decreases with density (0.9995, 0.9982 and 0.9967 over terciles of prerequisites per issue), as RQ1 predicts, but real sprints sit at the sparse end. The whole room for improvement, measured by the exact optimum, is 1.19 [0.27, 2.11] thousandths, against 21.6 on synthetic graphs at $\lambda=1$ (Figure~\ref{fig:svr}).

\emph{Unlock bonuses tuned on synthetic graphs do not transfer.} L1 and L2 are indistinguishable from L0 on average ($-0.08$ and $-0.60$ thousandths) and worse than L0 in 53\% and 59\% of sprints. L3 gains $+0.73$ [$-0.14$, 1.61] but is worse in 44\% of sprints. Smaller bonuses work better on real data (L1 with $\theta=0.1$: $+0.76$ [0.16, 1.36], worse in 3.5\% of sprints), so the useful strength depends on dependency density and cannot be fixed once from simulation.

\emph{Rollout is non-worsening on the declared surrogate.} L4 gains $+0.77$ [0.10, 1.43] thousandths, improves on L0 in 30\% of sprints and is never worse on $J$, as Proposition~\ref{prop:rollout} guarantees. Its runtime was 0.03~s per sprint (median, maximum 3.9~s). ODP + local search and Sidney (ODP in modules) reach the exact optimum in 99\% and 100\% of sprints. With $\beta=0.3$ or $\gamma=0$, L0 reaches 0.9976 and 0.9980, L4 gains $+1.03$ and $+0.86$ thousandths and is never worse, and the exact optimum gains $+1.76$ and $+1.32$: the conclusions are unchanged.

\emph{Historical issue-resolution order differs from the evaluated orders.} Ordering issues by recorded resolution timestamp reaches 0.929 [0.924, 0.935] of the surrogate optimum, is worse than L0 in 94\% of sprints, and respects 76\% of the recorded dependencies (all of them in 63\% of sprints). Resolution timestamps record closing, not starting, and priority is only a proxy for value. Moreover, this retrospective order violates some recorded precedence links and is therefore not a feasible comparator to the constrained optimum. It is not a measure of team performance. This gap motivates studying adoption of an explicit, dependency-valid order, but closing timestamps cannot establish that such adoption would improve team performance.

\subsection{RQ3: surrogate gains do not reach calendar time}\label{sec:rq3}

\begin{table}[t]
\centering\small
\caption{Synthetic calendar criterion ($n=25$, $\lambda=1$, durations $E\,e^{\varepsilon}$, $\varepsilon\sim N(0,0.35^2)$, not seen by the policies): value-weighted completion relative to the best evaluated order (means over 30 seeds).}\label{tab:calsynth}
\begin{tabular}{lcc}
\toprule
Policy & 1 team & 3 teams\\
\midrule
L0 native ODP & 0.908 & 0.915 \\
$\PVS_2$, eligible greedy & 0.877 & 0.901 \\
Value/effort (WSJF) & 0.972 & 0.941 \\
L1 count ($\theta=1$) & 0.945 & 0.955 \\
L2 value ($\theta=0.25$) & 0.947 & 0.962 \\
L3 multi-hop ($\theta=0.1$) & 0.943 & 0.964 \\
L4 one-step rollout & 0.944 & 0.970 \\
ODP + local search & 0.954 & 0.971 \\
Sidney (ODP in modules) & 0.959 & 0.972 \\
Exact optimum & 0.961 & 0.974 \\
Random topological & 0.760 & 0.833 \\ 
\bottomrule
\end{tabular}
\end{table}

In our retrospective real-sprint calendar replay, the exact surrogate optimum is no better than L0 in the modeled completion criterion (one team: 0.954 versus 0.955, three teams: 0.977 versus 0.975), and the anticipation levels are slightly worse on one team (0.938--0.952, Table~\ref{tab:tawos}). The best order in calendar time is value-to-effort (0.970 and 0.983), because recorded working time follows story points more closely than the composite ERS. On dense synthetic graphs (Table~\ref{tab:calsynth}) anticipation does carry over to calendar time: on three teams, L4 reaches 0.970 against 0.915 for L0 (paired gain 54.5 [47.7, 61.3] thousandths). On one team, value-to-effort is again best (0.972), ahead of the exact surrogate optimum (0.961). In these evaluations, a gain in $J$ transfers to earlier modeled completion mainly when dependencies are dense and ERS tracks duration. Neither condition is well supported by the observed sprint proxies.

\subsection{RQ4: governance checks}\label{sec:rq4}
\emph{Pinned exponent and rounding.} On the 960 synthetic instances, the contract gives the same complete order under $\alpha_w=1$ and $\alpha_w=2$ in 955 cases (99.5\%), and the rounded order equals the unrounded one in 99.6\%. Real scores are integer-valued and ties are common: at 64\% of ODP steps on real sprints, two or more eligible issues share the smallest rounded distance. Yet $\alpha_w$ changes the order in only 4 of 373 sprints, without changing $J$, because tied issues usually have equal BVS and ERS and the name decides. The exponent is thus a governed parameter with a small but non-zero operational effect, and the name tie-break matters more than its rarity in continuous simulations suggests.

\begin{table}[t]
\centering\small
\caption{Surrogate gate on real sprints: alternatives kept because they strictly lower $J$, number with calendar data, and calendar outcome against L0 (worse / better, mean change in thousandths of the best calendar value) on 1 and 3 teams.}\label{tab:gate}
\begin{tabular}{lcccccc}
\toprule
& & & \multicolumn{2}{c}{1 team} & \multicolumn{2}{c}{3 teams}\\
\cmidrule(lr){4-5}\cmidrule(lr){6-7}
Alternative & kept & with data & worse / better & $\Delta$ & worse / better & $\Delta$\\
\midrule
L1 count ($\theta=1$) & 61 & 39 & 21 / 18 & -26.4 & 18 / 20 & -4.7 \\
L2 value ($\theta=0.25$) & 51 & 29 & 16 / 13 & -17.1 & 12 / 16 & +3.8 \\
L3 multi-hop ($\theta=0.1$) & 80 & 56 & 25 / 31 & -18.5 & 20 / 34 & +0.8 \\
L4 one-step rollout & 111 & 86 & 38 / 48 & -10.7 & 32 / 51 & +4.8 \\
ODP + local search & 116 & 90 & 33 / 57 & +6.8 & 26 / 62 & +14.2 \\
Exact optimum & 116 & 90 & 31 / 59 & -2.1 & 27 / 60 & +7.9 \\ 
\bottomrule
\end{tabular}
\end{table}

\emph{A surrogate gate is not a safety guarantee.} Keeping an alternative only when it lowers $J$ removes every surrogate loss by construction, but not calendar losses (Table~\ref{tab:gate}). On one team, kept alternatives are worse than L0 in calendar time in 34\% (exact optimum) to 54\% (L1) of sprints with data. For the rollout, this occurs in 38 of 86. The gate is useful to exclude alternatives that are worse on the declared objective. It cannot certify operational benefit.

\emph{BA-W1.} This comparison concerns the fixed Blackridge Atlas Wave 1 acceptance configuration, not observed bank-project delivery. All reported candidate sequences satisfy the 18 declared precedence arcs among its 14 Feature Blocks. This is an algorithmic feasibility check, not evidence that work was started in the recommended order or that interruptions were prevented. On the acceptance graph, L0 and L4 give $J=3711$ and ODP + local search and Sidney (ODP in modules) reach the optimum 3705 (Appendix~\ref{app:ba}). ODP places block 4.3 before 2.1, whereas the optimum advances the chain $2.1\to2.2\to2.3\to3.1$ first: a concrete instance of downstream blindness with a regret of 6 units (0.16\%). The rollout does not find it, because a single alternative first step completed greedily does not discover a multi-step chain. The unlock levels do worse than L0 here (L2 with $\theta=0.25$: 3717, L3: 3721--3730), a reminder that a bonus can over-promote prerequisites even when the plain ratio order is better.

\section{Discussion}
\subsection{How much anticipation is useful?}
Anticipation pays when dependencies are dense: on synthetic graphs with about one prerequisite per block, L0 leaves 2--6\% of the surrogate on the table and every level recovers most of it. On the real sprints studied, dependencies are sparse and L0 is within 0.2\% of the optimum. The unlock bonuses L1--L3 then become a liability: their best strength depends on density, and a value tuned on one graph family is wrong on another. The rollout has no such parameter and cannot be worse than L0 on $J$. It is a defensible default comparison for $J$, but it carries no guarantee of operational benefit. Where an exact or near-exact answer is cheap, as it is on every sprint studied here (Sidney with module DP, or local search), it should be shown next to L0 so that the distance to the optimum is visible in the Wave record.

\subsection{What the surrogate does not measure}
\emph{Scope of the empirical analogy.} The sprint-derived dataset tests the scheduling subproblem, not the D-POAF framework as a whole. A governed Wave includes more than an ordered set of work items. In particular, it carries context and decision authority across four lifecycle macro-phases, with traceable PoD, PoV and PoR expectations~\cite{dpoaf_spec,dpoaf_terms}. Jira priorities, story points and recorded internal links do not reconstruct those obligations. The reported 0.9981 ratio is therefore proximity to an optimum \emph{after proxy construction}, not measured Wave performance or evidence of D-POAF conformance.

BVS is a planning estimate. ERS is a composite score, not elapsed time. The real-sprint calendar results show that an order optimal for $J$ is not necessarily earlier in calendar time, and that value-to-effort can be better when working time follows the effort estimate. The evaluation should distinguish technical completion, validated delivery, release time and realized post-release value. If value is only realized at a Wave release gate, earlier technical completion is not earlier value.

\subsection{Operational interpretation: prerequisite readiness and avoidable disruption}
A precedence-valid plan addresses a concrete coordination problem beyond the numerical surrogate. If Feature Block $b$ declares prerequisite $a$, an ODP order does not place $b$ before $a$. Making that constraint visible before execution can support planning discussions about what is ready to start and why another block must be completed first. The property holds for the \emph{declared} graph and the \emph{recommended} order. It does not identify missing, late-discovered or incorrectly modeled dependencies, and it cannot ensure that an executing team follows the order. Discovery and review of prerequisite links belong upstream in Shape \& Align and remain necessary throughout a Wave.

Interruption and work-fragmentation studies suggest why prerequisite readiness may matter for developers, although they do not evaluate ODP~\cite{sanchez2015fragmentation,ma2024breaking}. A team that begins work before a prerequisite is usable may have to pause, switch tasks, wait, or revise completed work. These are \emph{plausible mechanisms}, not benefits measured here. Our results establish feasibility under known arcs and quantify performance on $J$. They do not measure dependency-induced pauses, context-switch counts, rework effort, planning effort, or a causal effect on delivery time. In particular, the historical Jira resolution timestamps cannot reconstruct when blocked work was first attempted.

A separate observational and prospective evaluation should distinguish \emph{dependency-known-at-planning-time} from links recorded later, and log attempted starts, actual starts, prerequisite readiness, pause reasons, resumptions and rework. This would permit direct estimation of avoidable dependency-related disruptions without treating a score-based optimum as evidence of operational savings. Such evaluation should be performed on completed, governed Waves when available, with public issue-tracking histories used only where their timestamps and links support the proposed measures.

\subsection{Toward a governed comparison, not an automatic switch}
The safest product design keeps L0 as the default and computes alternatives as read-only candidates against the same immutable snapshot. For every proposal, the Wave record should preserve the snapshot hash, scoring policy, $\alpha_w$ and rounding precisions. It should also record the target closure, prerequisite arcs, eligible-set decisions and ordered block identifiers. For each alternative, it should record the algorithm and its parameters (for L1--L3, $\theta$ and the graph density at which it was chosen), implementation version, runtime and replay identifier. Finally, the record should include $J$, the L0 value, the exact optimum when available, and explicit approval or rejection with a human rationale. No realized effort or reward should enter a retrospective snapshot, which would leak future outcomes into the recommendation.

\subsection{Testable adoption criteria}
Before a non-default level is offered in production we propose falsifiable requirements. \emph{Conformance:} for the same frozen snapshot, L0 agrees bit for bit with the reference service, including closure, cycle refusal, PVS tie-break, rounding and names. Every alternative is dependency-valid. \emph{Reproducibility:} every result replays from the stored policy and snapshot. \emph{Comparative quality:} the exact optimum or a certified bound is shown where computable, and gains and losses are reported with paired uncertainty at the Wave level. \emph{Operational outcomes:} independent project data distinguish projected score burden from dependency-induced waiting, interrupted starts, resumptions, rework, recorded working time, capacity, delivery dates and post-release value. User studies cover comprehension and overriding. \emph{Governance:} an approved non-default level has an owner, a scope, a rollback path and monitoring. None of these conditions is claimed to be met by the present study.

\section{Threats to Validity}
\textbf{Construct.} A sprint and a D-POAF Wave have different semantics. The former is a timeboxed Scrum event with goals and an adaptable backlog, while the latter is a verifiable-progress unit and governance container with scope, proof obligations and lifecycle-wide decision records. A Jira issue does not necessarily correspond to a Feature Block. Jira priority is a proxy for business value, story points are relative estimates rather than elapsed effort, and constant risk is assumed rather than observed. These proxies cannot evaluate PoD, PoV, PoR, Wave governance or organizational adoption. Time ``In Progress'' is a noisy working-time signal that misses waiting and parallel work. The modeled calendar criterion credits each proxy item at completion.
\textbf{Internal.} Jira links may be incomplete, and within-sprint filtering omits prerequisites outside the sampled issue set, so observed graphs may be considerably sparser than the underlying project dependencies. Resolution order is not start order. Parameters of L1--L3 were selected on synthetic data and then applied to real data without tuning, which is the intended test but also the reason for their weak transfer.
\textbf{External.} Twenty-one projects from one public dataset, one of which contributes 152 sprints, do not represent industrial practice. Project-level intervals mitigate but do not remove this clustering. At $n=100$, 11 of 150 synthetic instances are excluded because an exact module exceeds the computational cap, potentially selecting easier cases. The four largest sprints are excluded from L4 comparisons. BA-W1 is one project-specific private-beta acceptance case, but its scores and dependency configuration were used as controlled testing inputs. It is not a completed client-bank deployment, and the private-beta scenario brief does not itself establish achieved operational outcomes. Synthetic benchmark results separately describe the declared random generator.
The private-beta scenario is defined by Inovionix and the numerical BA-ODP-01 inputs were configured for acceptance checking. The scenario brief is not independent evidence of completed development, and reviewer inspection of non-public project artifacts is subject to authorization.
\textbf{Conflict of interest.} The authors develop D-POAF (see Competing Interests). Every comparison includes methods from outside D-POAF that outperform the native rule. The separate research archive is intended to support computational reproducibility without disclosing production source code.

A prospective study should predeclare primary operational outcomes (dependency-related blocked starts, waiting and rework, independently observed value-weighted completion, release gates and value recorded after release), freeze the Matrix and scoring policy before recommendations, and timestamp when prerequisite links became known. It should run alternatives in shadow mode before any controlled adoption, and report adverse-case rates, explainability, computational cost and human acceptance alongside means and intervals. Observational comparisons would still require explicit handling of project-level confounding.

\section{Conclusion}
Explicitly rewarding newly unlocked Feature Blocks reduces the gap left by greedy ODP on dense synthetic graphs. On sparse dependency graphs reconstructed from historical Jira sprints, the native rule is already within 0.2\% of the exact surrogate optimum, tuned unlock bonuses do not transfer and surrogate gains do not reliably translate into earlier modeled calendar completion. These sprint-derived results test the transfer of sequencing rules, not equivalence between a sprint and a governed D-POAF Wave. The one-step rollout is non-worsening by construction for $J$ only, and exact references were computationally feasible on the sprint-derived instances studied. A dependency-valid recommendation can also make prerequisite readiness explicit before implementation, potentially helping teams avoid starts that are infeasible under their known dependency graph. Whether that reduces waiting, task switching or rework is a distinct empirical question. The supported design is not to abandon ODP but to keep a verifiable Level~0, show the rollout and the exact optimum as auditable read-only comparisons, and reserve unlock bonuses for contexts whose dependency density has been measured.

\Needspace{13\baselineskip}
\section*{Availability and Reproducibility}
The public aggregate benchmark results are published on Zenodo (version 1.0) under DOI \doi{10.5281/zenodo.23762369}~\cite{packageB2026}. The public package contains aggregate and synthetic results, benchmark tables and provenance notes without the proprietary platform source code or the detailed BA-W1 configuration. A separate restricted verification package has been prepared with standalone research scripts, exact-solver tests, derived TAWOS scheduling instances, the BA-ODP-01 scoring and dependency configuration, and experimental outputs. These research scripts are independent of the D-POAF SaaS implementation. Qualified reviewers may request restricted materials from the corresponding author (\href{mailto:sara.ihsine@inovionix.com}{sara.ihsine@inovionix.com}). For anonymous peer review, requests can instead be routed through the journal editor. Access is subject to rights-holder approval, third-party data terms and any necessary confidentiality arrangements. Not every component can necessarily be released, and a supervised or redacted verification may be offered where sharing is not authorized. No production D-POAF source code is available. A local reproduction command (\texttt{bash run\_all.sh}) is supplied in the restricted package. Experiments used Python~3.13, NumPy~2.5.3, SciPy~1.18.1, pandas~3.0.5 and NetworkX~3.6.1~\cite{hagberg2008networkx}. The restricted verification materials have no public DOI assigned in this manuscript.

\section*{Competing Interests}
Azzeddine Ihsine and Sara Ihsine contribute to the D-POAF framework and are affiliated with Inovionix, which develops a software platform implementing parts of the framework. The authors declare no other competing interests.

\bibliographystyle{unsrtnat_hyphenated}
\bibliography{refs}
\appendix
\section{Parameters}\label{app:params}
\begin{table}[H]
\centering\small
\begin{tabular}{p{0.2\linewidth}p{0.72\linewidth}}
\toprule
Synthetic & Master seed 20260827, 30 seeds, $\BVS\sim N(5.5,2^2)$ truncated to $[1,10]$, effort lognormal (median 5, log-s.d.\ 0.7, truncated to $[1,40]$), risk $U[1,10]$, $\ERS=1+0.1E+0.1R$, prerequisites $\min(\mathrm{Poisson}(\lambda),3)$ among earlier blocks, $n=12$: $\lambda\in\{0,0.5,1,1.5\}$, eight instances per seed and level, keys shared with the companion article, scaling: $n\in\{25,50,100\}$, $\lambda=1$, five instances per seed, calendar: $n=25$, $\lambda=1$, eight instances per seed, duration noise log-s.d.\ 0.35, teams $\{1,3\}$.\\
Levels & $\theta\in\{0.1,0.25,0.5,1\}$ selected on seeds 0--14 at $\lambda=1$ (L1: 1, L2: 0.25, L3: 0.1), L3 discount $\delta=0.5$, L0 tie-break $\alpha_w=2$ then name, local search: first-improvement insertion, at most 50 passes.\\
Exact solver & Sidney decomposition into inclusion-minimal ratio-maximal initial sets (Dinkelbach + minimum cut, NetworkX), exact subset DP per module, module limit 20 (synthetic) and 22 (real).\\
TAWOS & Links ``blocks'' (Blocks, Blocker, Blockers) and ``depends on'' (Depends, Dependency) inside a sprint, excluded resolutions: Duplicate, Won't Fix, Cannot Reproduce, Invalid, Incomplete, Not A Bug, Won't Do, Abandoned, Obsolete, Rejected, Declined, priority levels: Trivial, Trivial-P5, Lowest $=1$, Minor, Minor-P4, Low $=2$, Major, Major-P3, Medium, To be reviewed (and missing) $=3$, Critical, Critical-P2, High $=4$, Blocker, Blocker-P1, Highest $=5$, $\BVS=2\times$level, story points $\le40$, else sprint median, $\ERS=1+0.1E+0.1\times5$ (sensitivities $\beta=0.3$, $\gamma=0$), recorded durations: TAWOS \texttt{In\_Progress\_Minutes} $>0$, rollout skipped above 120 issues.\\
\bottomrule
\end{tabular}
\end{table}

\section{BA-W1: Blackridge Atlas Private-Beta Wave 1 Acceptance Case}\label{app:ba}
\begin{table}[H]
\centering\small
\caption{BA-W1 acceptance configuration for Wave 1 of the Blackridge Atlas private-beta risk-scoring API (14 Feature Blocks, 18 dependency links, $\alpha=2$): objective $J$ (lower is better) and $J^\star/J$. All orders are dependency-valid. This is one controlled private-beta case, not a population of independent projects, so no statistical interval is meaningful.}
\begin{tabular}{lcc}
\toprule
Policy & $J$ & $J^\star/J$\\
\midrule
Exact optimum & 3705 & 100.00\% \\
L0 native ODP & 3711 & 99.84\% \\
$\PVS_2$, eligible greedy & 3792 & 97.71\% \\
L1 count ($\theta=0.25$) & 3711 & 99.84\% \\
L1 count ($\theta=1$) & 3718 & 99.65\% \\
L2 value ($\theta=0.25$) & 3717 & 99.68\% \\
L3 multi-hop ($\theta=0.1$) & 3721 & 99.57\% \\
L3 multi-hop ($\theta=0.25$) & 3730 & 99.33\% \\
L4 one-step rollout & 3711 & 99.84\% \\
ODP + local search & 3705 & 100.00\% \\
Sidney (ODP in modules) & 3705 & 100.00\% \\ 
\bottomrule
\end{tabular}
\end{table}

\end{document}